\documentclass[12pt, preprint]{aastex}

\def\deg{\ifmmode^\circ\else$^\circ$\fi}

\def\arcs{\ifmmode {''}\else $''$\fi}
\def\arcm{\ifmmode {'}\else $'$\fi}
\def\parcs{\sa=.07em \sb=.03em
     \ifmmode $\rlap{.}$^{\scriptscriptstyle\prime\kern -\sb\prime}$\kern -\sa$
     \else \rlap{.}$^{\scriptscriptstyle\prime\kern -\sb\prime}$\kern -\sa\fi}
\def\parcm{\sa=.08em \sb=.03em
     \ifmmode $\rlap{.}\kern\sa$^{\scriptscriptstyle\prime}$\kern-\sb$
     \else \rlap{.}\kern\sa$^{\scriptscriptstyle\prime}$\kern-\sb\fi}

\def\Msun{\mbox{M$_{\odot}$}}

\def\han {\mbox{{\rm H}$\alpha$}}
\def\ha{\han}
\def\hb{\mbox{{\rm H}$\beta$}}

\def\h0{\mbox{H$_0$}}

\def\spose#1{\hbox to 0pt{#1\hss}}
\def\simlt{\mathrel{\spose{\lower 3pt\hbox{$\mathchar"218$}}
     \raise 2.0pt\hbox{$\mathchar"13C$}}}
\def\simgt{\mathrel{\spose{\lower 3pt\hbox{$\mathchar"218$}}
     \raise 2.0pt\hbox{$\mathchar"13E$}}}
\def\lsim{\rlap{$<$}{\lower 1.0ex\hbox{$\sim$}}}
\def\gsim{\rlap{$>$}{\lower 1.0ex\hbox{$\sim$}}}

\received{2002 July 30}
\begin{document}

\title{Emission Line Galaxies in the STIS Parallel Survey II: Star Formation Density\footnotemark}

\footnotetext[1]{Based on observations made with the NASA/ESA {\em
Hubble Space Telescope}, obtained from the data archive at the
Space Telescope Science Institute, which is operated by the
Association of Universities for Research in Astronomy, Inc., under
NASA contract NAS 5-26555.}

\author{
Harry I. Teplitz,\altaffilmark{2,}\altaffilmark{3} Nicholas R.
Collins,\altaffilmark{4} Jonathan P. Gardner, Robert S.
Hill,\altaffilmark{4} \& Jason Rhodes\altaffilmark{5} }

\affil{Laboratory for Astronomy and Solar Physics, Code 681,
Goddard Space Flight Center, Greenbelt MD 20771 \\Electronic mail:
hit@ipac.caltech.edu}

\altaffiltext{2}{CUA Research Associate}

\altaffiltext{3}{New affiliation -- SIRTF Science Center}

\altaffiltext{4}{SSAI}

\altaffiltext{5}{NRC Fellow}

\begin{abstract}
  
  We present the luminosity function of [OII]-emitting galaxies at a
  median redshift of $z=0.9$, as measured in the deep spectroscopic
  data in the STIS Parallel Survey (SPS).  The luminosity function
  shows strong evolution from the local value, as expected.  By using
  random lines of sight, the SPS measurement complements previous deep
  single field studies.  We calculate the density of inferred star
  formation at this redshift by converting from [OII] to \ha\ line
  flux as a function of absolute magnitude and find
  $\dot{\rho}=0.043\pm 0.014$\ \Msun\ yr$^{-1}$\ Mpc$^{-3}$\ at a
  median redshift $z\sim 0.9$\ within the range $0.46<z<1.415$ ($H_0 =
  70$ km s$^{-1}$\ Mpc$^{-1}$, $\Omega_M=0.3, \Omega_{\Lambda}=0.7$).
  This density is consistent with a $(1+z)^4$\ evolution in global
  star formation since $z\sim 1$.  To reconcile the density with
  similar measurements made by surveys targeting \ha\ may require
  substantial extinction correction.

\end{abstract}


\section{Introduction}

The comoving density of global star formation in the Universe has
decreased significantly since a redshift of $z\sim 1-2$ (Hogg 2002;
Madau, Pozzetti, \& Dickinson 1998; Madau et al.\ 1996), but the
details of the evolution are still uncertain.  Extensive measurements
of the star-formation history have concentrated on surveys for \ha\ 
emission (e.g.\ Gallego et al.\ 1995, Yan et al.\ 1999, Tresse et al.\ 
2002) and the UV continuum (e.g.\ Lilly et al.\ 1996, Connolly et al.\ 
1997).  A correction factor is needed reconcile the two methods
(Calzetti et al.\ 1997; Tresse \& Maddox 1998; Glazebrook et al.\ 
1999) as dust obscuration is expected to hide some significant
fraction of star formation in the distant Universe.  Complementary
optical data is provided by targeting the [OII] $\lambda 3727$\ 
emission line, which can also provide a direct measurement of the rate
of star-formation, though its calibration depends on the metallicity
and ionization state of the gas, as well as the reddening (Kennicutt
1998).  Indeed, surveys at higher redshift will need to depend on the
oxygen line. At $z>4$, the \hb\ line redshifts out of the K-band,
making it inaccessible to ground-based NIR spectrometers.  As a
result, [OII] emission is the best rest-frame optical tracer of
star-formation available at high redshift until NGST.

In this paper, we present analysis of the deepest spectroscopic fields
in the STIS parallel survey (SPS; Gardner et al.\ 1998).  Our
motivation is to determine the comoving density of star-formation at
$z\sim 0.9$\ as inferred from the [OII] emission line.  The deep SPS
data cover 141 square arcminutes, and detect objects down to continuum
magnitudes of $m_{50CCD}\sim 26$\ in filterless imaging, with emission
line strengths as faint as $7.5 \times 10^{-17}$\ ergs cm$^{-2}$\ 
s$^{-1}$\ (Teplitz et al.\ 2002, hereafter Paper I).  These data are
complementary to other surveys of the [OII] luminosity density.  Hogg
et al.\ (1999a; hereafter H99) measured [OII] emission in 375 galaxies
with $R<23.3$\ from the redshift survey data in the 50 square
arcminute ``Caltech area" (Cohen et al.\ 2000) around the Hubble Deep
Field North (HDF; Williams et al.\ 1996).  The HDF spectra obtained
with LRIS (Oke et al.\ 1995) on the Keck telescopes measure emission
lines to a limit 15 times fainter than the SPS.  Prior to H99, the
best measurement of the [OII] luminosity density was obtained the
Canada France Redshift Survey (CFRS; Lilly et al.\ 1995) by Hammer et
al.\ (1997, hereafter CFRS9).  A significant increase in [OII]
luminosity is seen with lookback time in the range $0< z< 1$\ (Cowie
et al.\ 1996; Gallego et al.\ 2002).

Throughout the paper, we will adopt a flat, $\Lambda$-dominated
universe ($H_0 = 70$ km s$^{-1}$\ Mpc$^{-1}$, $\Omega_M=0.3,
\Omega_{\Lambda}=0.7$), except where otherwise noted.

\section{The Data}

The slitless spectroscopic capability of STIS, using the G750L grating
with spectral resolution of $\sim 750$\ at 7500\AA, allowed a
random-field redshift survey for star-forming galaxies at $0.46 < z <
1.7$.  The deepest 219 of the $\sim 2500$\ fields, which were observed
by STIS in parallel with targeted HST observations, contain 78 objects
identified with [OII] emission (Paper I).  Of these, 14 have multiple
emission lines, providing secure redshifts.  For the rest, the
identification of a single line as [OII] is based on the non-detection
of other lines, together with photometric information from the paired
direct filterless imaging.  A signal to noise limit of $\sigma=2.5$\ 
was imposed for the strongest line in an object in order for it to be
included in the catalog.  For the present analysis, we impose a
redshift cutoff of $z \le 1.415$\ to avoid the objects in the poor
sensitivity region at $\lambda_{obs}>9000$\ \AA.  In total, we include
71 of the 78 [OII]-emitters from Paper I.

\begin{figure}
\plotone{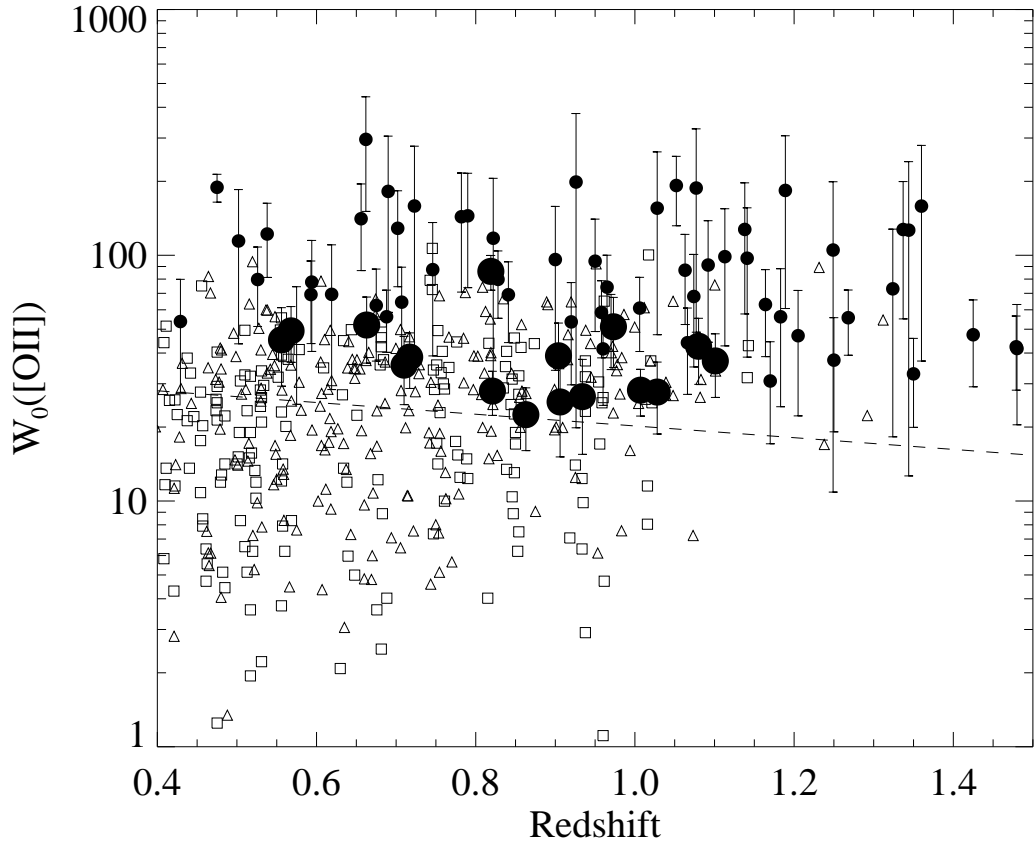}
\caption{The redshift distribution of rest-frame [OII] equivalent widths in the SPS.  The small filled
circles are SPS data with relatively poor continuum spectra ($SNR<1$\ in two pixels), and the large
filled circles have good continuum detections ($SNR>1$ in two pixels).  The measurements in the
HDF (open squares; H99) and the CFRS (open triangles; CFRS9) are shown for comparison.  The dashed line
indicates the $W_{obs}>35$\ \AA\ cutoff of the SPS data.
\label{fig: ew vs z} }
\end{figure}

The SPS detections are biased towards objects with a strong contrast
between line and continuum.  This bias means most SPS line-emitters
are morphologically compact and that most of the emission lines have
high equivalent width (EW).  The SPS detection limit is $\sim 35$\ 
\AA\ for observed EW, consistent with the values for most galaxies
undergoing rapid star-formation (Kennicutt 1992).  The measured EW
value is uncertain for most SPS objects due to low significance of
continuum measurement in the spectra.  Typically the continua are
measured with a signal-to-noise ratio (SNR) of $\sigma_c \sim 0.5$\ 
per two-pixel resolution element.  The continuum is measured over many
resolution elements (typically $20-30$), but is still a major source
of error in the EW.  Figure \ref{fig: ew vs z}\ compares the EW
distribution for the SPS with that of other [OII] surveys at similar
redshifts.  The SPS objects have a median rest-frame EW for [OII] of
$W_0(OII)\sim 70$\ \AA, but many of the $W_0(OII)>100$\ \AA\ objects
have poor continuum detections, so that the EW may be overestimated
(see H99).  It appears from the figure that the SPS is missing a
substantial portion of low EW lines, particularly at low redshift.  At
high redshift, low EW lines may simply have been missed in all of the
surveys.  Nonetheless, CFRS9 and Cowie et al.\ (1996) both conclude
that there is a strong increase in the fraction of galaxies with
strong [OII] emission ($EW([OII])\simgt 25$\ \AA) at $z>0.5$.

\subsection{Incompleteness}

\begin{figure}[ht]
\plotone{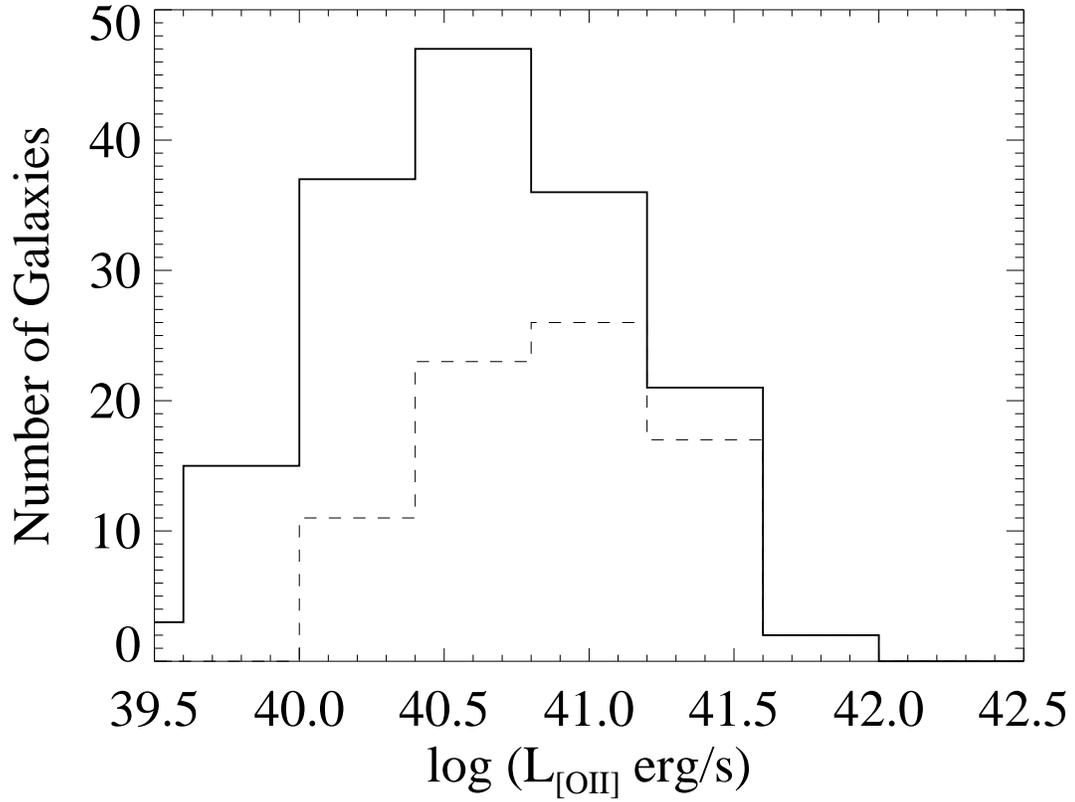}
\caption{The histogram of [OII] luminosities in the HDF from H99 ($H_0 = 100$\ km s$^{-1}$\ Mpc$^{-1}$, 
$\Omega_M=0.3, \Omega_{\Lambda}=0$).  The solid line is the 
entire sample, and the dashed line is the fraction of the sample with observed EW$>35$\AA;
i.e., those EW greater than the 
SPS equivalent width cutoff. \label{fig: ew_check}}
\end{figure}

The SPS survey detects $W(OII)>35$\ \AA\ in the observational frame.
Figure \ref{fig: ew_check} shows the distribution of missing low-EW
objects as a function of [OII] luminosity, based on H99.  In the
analysis of the SPS sample, we correct for this incompleteness.
Because the H99 published data only present the equivalent width
distribution for objects with continuum detection ($2\sigma$), this
incompleteness correction is strictly an upper limit.

The primary source of incompleteness in the SPS sample is missing
objects that would have met the selection criteria.  Visual detection
by multiple observers was more accurate than any available automated
detection scheme, but human error remains a factor.  Simulated
spectroscopic images were used to determine the completeness level of
our detection scheme.  Test images were created by inserting known
spectra into SPS frames and trying to recover them.  Two spectra were
used as the basis of this experiment, one with $W(OII)\sim 100$\ and
the other with $W(OII)\sim 35$.  Both spectra were high signal to
noise ($\sigma_{OII}>8$), but were scaled to lower significance in the
test frames.  The emission line itself was moved within the spectrum
to prevent identification by pattern matching.  The spectra were
inserted at random spots in randomly chosen images and scaled to an
emission line signal-to-noise ratio of 3--7.  For each test case, 300
simulated frames were created.  In some frames, continuum spectra were
also added without an emission line to further disguise the simulated
data.  Table \ref{table: incompleteness}\ lists the resulting
incompleteness as a function of EW and SNR.  The completeness of the
survey at any given SNR and EW value is obtained by interpolating
between the data points in the table.

\section{Calculating the Luminosity Function}

We use the $1/V_{\mbox{max}}$\ method to calculate the luminosity
function (LF) of [OII] emission lines.  For each galaxy we determine
the comoving volume over which it could be detected in our survey.
The volume is different for each of the parallel fields, based on
the limiting flux. Assuming we have $j$\ detections in $k$\
fields, then for a single galaxy in a single field we have,
following Hogg (1999b) :
\begin{equation}
V_{\mbox{max},j,k} =
    \biggl( \frac{c}{H_0} \biggr) \int \int_{z_{min}}^{z_{max}}{\frac{(1+z)^2 D_A^2}{E(z)} d\Omega dz }
\end{equation}

where,

\begin{equation}
E(z) \equiv \sqrt{\Omega_M (1+z)^3 + \Omega_k (1+z)^2 + \Omega_{\Lambda}}
\end{equation}

and, $D_A$\ is the angular diameter distance, which in the case of a flat, matter dominated universe is:

\begin{equation}
D_A = \biggl( \frac{c}{H_0} \biggr) \frac{1}{(1+z)} \int_{0}^{z} \frac{dz^{\prime}}{E(z^{\prime})}
\end{equation}

Further, $d\Omega$\ is a function of z, because the available area
depends on the placement of the observed wavelength of the line and the position of
the object within the field of view.  That is, if
the object is too far to one side, the line will fall off the edge
of the detector.  So for an $\textrm{npix}_x \times \textrm{npix}_y$\ detector, a central
wavelength $\lambda_c$, a line at rest wavelength $\lambda_0$\,
and a dispersion $d$,

\begin{equation}
\int d\Omega = \biggl( \textrm{npix}_x - \frac{|\lambda_0 (1+z) - \lambda_c
|}{d} \biggr) \textrm{npix}_y
\end{equation}

The effective area of the detector is slightly smaller than the default field of view (51.2\arcs $\times 52.2$\arcs), 
since we only
consider the region of the registered frame with the full exposure
time. Similarly, we do not consider area blocked by very bright
stars or local galaxies (see the first table in Paper I).

The volume integral is over $z_{min}$\ to $z_{max}$, as defined by
the signal-to-noise in the line.  In an exposure with limiting
flux $f_{lim}$, a line that is detected at $\lambda_{obs}$\ with
flux $f_{obs}$\ and originating at $z_{obs}$\ would be have been
observable at any $z$\ such that

\begin{equation}
D_L (z) < D_L (z_{obs}) \sqrt{\frac{f_{obs}}{f_{lim} S(z)}}
\end{equation}
where $S(z)$\ is the sensitivity function of the spectrum, normalized to unity at $\lambda_{obs}$, and
$D_L$\ is the luminosity distance

\begin{equation}
f \equiv \frac{L}{4\pi D_L^2(z)}.
\end{equation}

\begin{figure}[ht]
\plotone{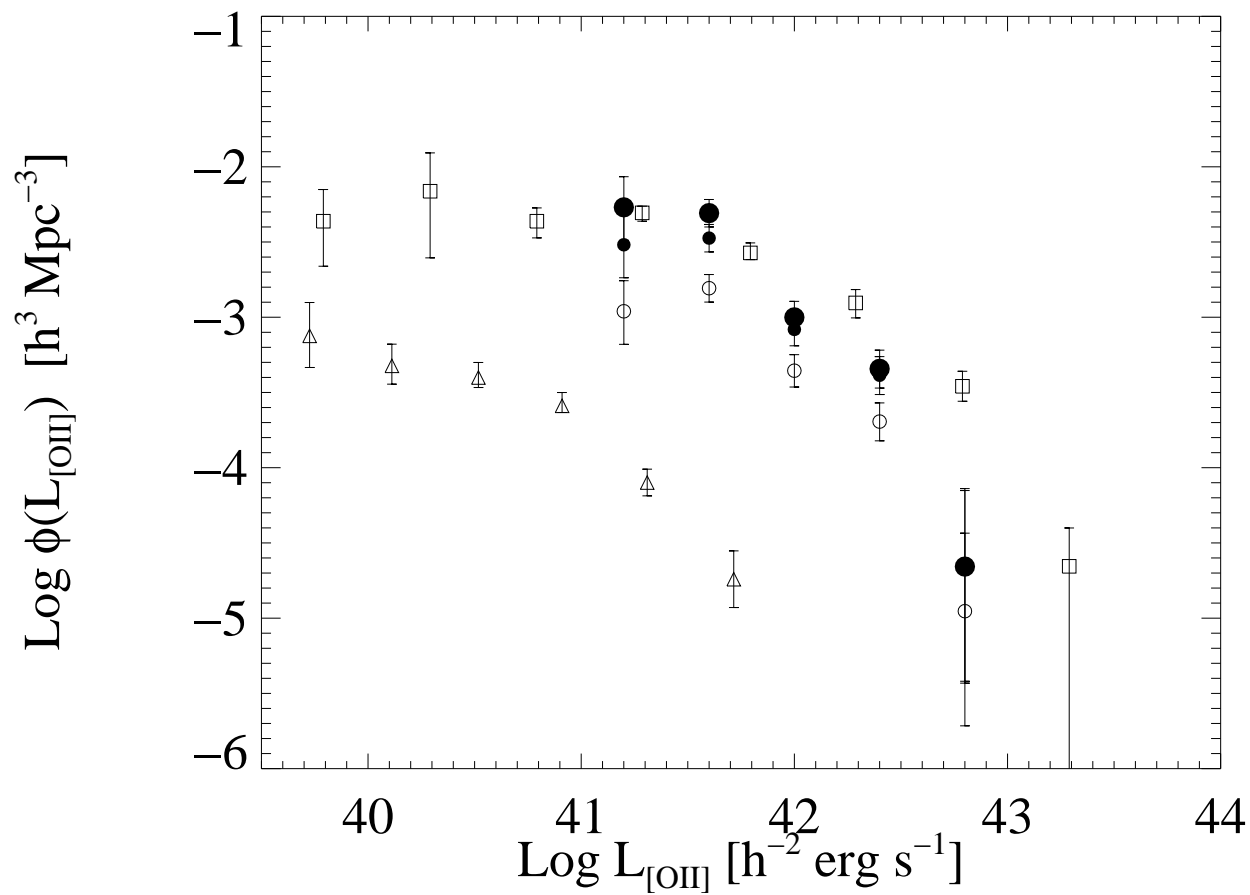}
\caption{The luminosity function of [OII] emission in the SPS compared to the $0.35<z<1.5$\ sample in the HDF (H99).
The open circles are the raw SPS detections.  Small filled circles have the incompleteness correction applied
down to the EW cutoff and the large filled circles have the additional correction for missing objects below
the EW limit (see text).  The open triangles are the [OII] LF for the local Universe (Gallego et al.\ 2002) and
the open squares are the HDF (from H99).
\label{fig: hogg_lf} }
\end{figure}

The individual $V_{\mbox{max},j,k}$\ for the each of the survey fields are summed for each galaxy
to obtain its $V_{\mbox{max},j}$.  Finally, the number density of galaxies in a luminosity 
bin of width $\Delta( \log{L})$\ is obtained by:

\begin{equation}
\phi = \frac{1}{\Delta (\log{L})} \eta \sum_j \frac{1}{V_{\mbox{max},j}}
\end{equation}
where $\eta$\ is the inverse of the completeness function at the
SNR and EW of the detected line.  
Figure \ref{fig: hogg_lf}\ shows the SPS [OII] luminosity function compared to
that of H99 and the local measurement (Gallego et al.\ 2002).  The SPS measurement appears
systematically underdense at high luminosities.  Only two objects contribute to the 
last $\log{L_{OII}}$\ bin, so missing or misidentified objects may be the cause of that 
bin being so extreme an outlier.  The flattening of the H99 LF at low luminosities compared to the
SPS measurement might be expected given the H99 magnitude limit, since H99 do not detect the high-EW
$R>23$\ galaxies that make up much of the SPS sample and may contribute to the faint end of
the LF.  The LF inferred from SPS
shows strong evolution with respect to the local LF, as expected.

\section{Comoving Density of Star-Formation}

\ha\ emission is a good indicator of star formation because it traces the ionizing 
flux from hot stars.  
To infer the star formation rate (SFR) in a galaxy from [OII] emission,
it is necessary to assume an [OII]:\ha\ ratio.  This ratio has an average of 0.45 for local 
galaxies (Kennicutt 1998), but is highly dependent on the metallicity and reddening of the
individual galaxy.  Jansen, Franx, \& Fabricant (2001) show that the ratio can vary by up to a 
factor of 7, and that it has a strong inverse correlation with continuum luminosity.  We adopt the 
relation they find for local galaxies with strong \ha\ emission (EW(\ha)$>10$\ \AA):
\begin{equation}
\log ( \frac{[OII]}{\ha} ) = 0.09 M_B + 1.42
\end{equation}
Jansen et al.\ find considerable scatter about this relation.
Apparently, the relation holds at higher redshift ($z\sim 1$; Tresse
et al.\ 2002), but it has not been tested on many objects, and it may
not hold in all cases (Hicks et al.\ 2002).  If we instead use the
average value for the ratio of [OII] to \ha, then we obtain similar
results though the distribution of \ha\ luminosities has less outliers;
that is, we find fewer galaxies with $L_{\ha} < 10^{42}$\ or
$>10^{43}$.

In order to apply the Jansen et al.\ relation, it is necessary to know
the rest-frame $M_B$\ of each galaxy.  The SPS images are taken with
the STIS filterless CCD, which admits a much wider wavelength range
than the B filter.  We adopt a continuum slope proportional to
wavelength, except in cases where a good detection of the continuum in
the spectrum allows direct measurement.  This conversion is an
important systematic error in the SPS data, and in the future could be
improved with photometry of the emission line objects in several
filters.  The average value of [OII]:\ha\ remains close to 0.45 for
the SPS sample using the Jansen et al.\ relation, but some individual
galaxies change by a factor of more than two.
  
\begin{figure}[ht]
\plotone{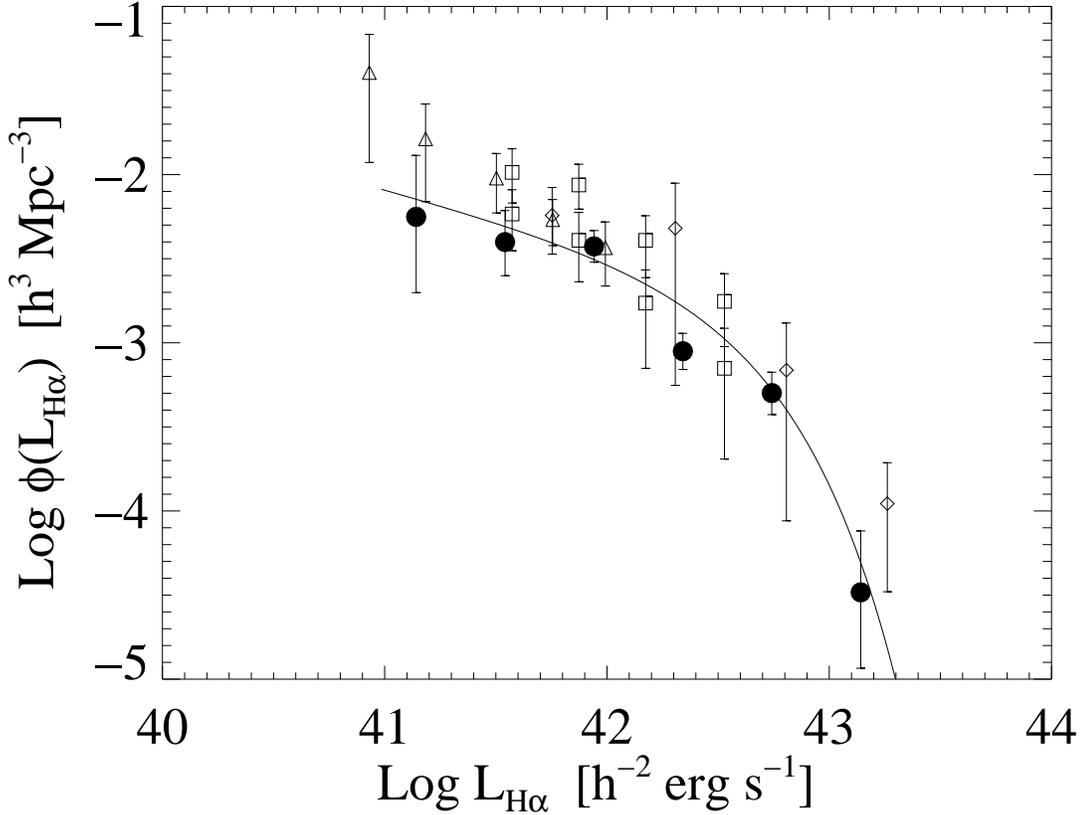}
\caption{The luminosity function for [OII] line fluxes converted to \ha\ (see text) compared to other measurements at
similar redshifts.
The large filled circles are the SPS values corrected for incompleteness including the missing low EW objects.  The
other data points are as follows --  open diamonds: $z\sim 1.3$\ from the NICMOS parallels(Yan et al.\ 1999); 
open squares: $z\sim 1.25$\ from NICMOS observations of the Groth strip
 (Hopkins et al.\ 2000), each bin plotted for both spectroscopically confirmed (lower) and all candidate (upper)
objects;  open upward triangles: $z\sim 0.73$\ from the CFRS (Tresse et al.\ 2002).
%; 
%open downward triangle: the local universe (Gallego et al.\ 1995).  
The solid line shows the Schechter function fit
with $\alpha=-1.35$. \label{fig: lf ha}}
\end{figure}

Figure \ref{fig: lf ha}\ shows the luminosity function of SPS emission
lines converted to \ha, compared to other measurements of LF(\ha) at
similar redshifts, from the literature.  Of particular interest are the
comparisons to measurements made with NICMOS slitless spectroscopy on
HST (Thompson et al.\ 1998).  The NICMOS parallels (McCarthy et al.
1999; Yan et al.  1999) are a survey comparable in many respects to
the SPS, since they are biased towards compact high-EW objects and
sensitive to a similar range of luminosities.  The NICMOS parallels do
not detect lines with $EW_{\mbox{rest}}(\ha)<50$\AA, but \ha\ EW is
typically 2.5 times that of [OII] (Kennicutt 1998).  
%The highest
%luminosity bin shows fair agreement between NICMOS and STIS, most
%likely because of the similarity in survey techniques.

The SPS measurement of the \ha\ LF appears systematically somewhat
lower than that of NICMOS.  The SPS objects are at a slightly lower
average redshift, and some evolution in the LF is expected.  In
addition, [OII] is subject to more extinction than \ha, though this
effect is mitigated by reddening in the galaxies used to fit the
Jansen et al. relation for [OII]:\ha.  
%The LF(\ha) inferred from SPS
%shows strong evolution with respect to the local LF, as expected.

We fit a Schechter function to our LF(\ha), even though the SPS fields do not contain enough 
detections to constrain
all three parameters ($\phi^*$,$L^*$, and $\alpha$) accurately.  We adopt the faint-end slope measured for the local
universe ($\alpha \sim -1.35$; Gallego et al.\ 1995), as do most of the \ha\ surveys.  
Hopkins, Connolly, \& Szalay (2000) use a steeper slope, $\alpha=-1.6$, in good agreement with 
that of Sullivan et al.\ (2000) for [OII].  If we had used this value, our final SFR densities would
have been $\sim 25$\% higher.
   The integral of the Schechter function is given analytically by
\begin{equation}
\phi^* L^* \Gamma(2+\alpha).
\end{equation}
Thus the integrated \ha\ luminosity density inferred from the SPS data is $5.9\times 10^{39}h_{70}$\ ergs s$^{-1}$\ at
a median redshift of $z\sim 0.9$.  This value is about 4.1 times the local \ha\ luminosity 
density (Gallego et al.\ 1995), but 
only half that measured
in the NICMOS parallels at $z\sim 1.3$\ (Yan et al.\ 1999). 

Assuming case B recombination, and a Salpeter initial mass function
truncated at 0.1 and 100 \Msun, the star formation rate in a galaxy can be inferred from the \ha\ luminosity
using the Kennicutt (1998) relation: 
\begin{equation}
SFR(\Msun\ yr^{-1}) = 7.9 \times 10^{-42} L(\ha)\ (\mbox{ergs s}^{-1}).
\end{equation}
Using this conversion, we obtain a comoving density of star formation
of $0.043\pm 0.014$\ \Msun\ yr$^{-1}$\ Mpc$^{-3}$\ at $z\sim 0.9$.
Figure \ref{fig: madau}\ compares the SFR density to values measured
from other rest-frame UV-optical surveys.  The SPS value is calculated
as an integral over a large redshift range ($0.46<z<1.415$), but it
can also be measured for a subset of the sample.  In Figure \ref{fig:
  madau}, we show the SPS density of star formation for the subsets of
the data in the ranges $0.46<z<1$\ and $1<z<1.41$, with median
redshifts of 0.72 and 1.14, respectively.  We do not plot the SFR
density inferred by H99, but these authors note that their SFR density
agrees with CFRS9.

\section{Discussion}

\begin{figure}[ht]
\plotone{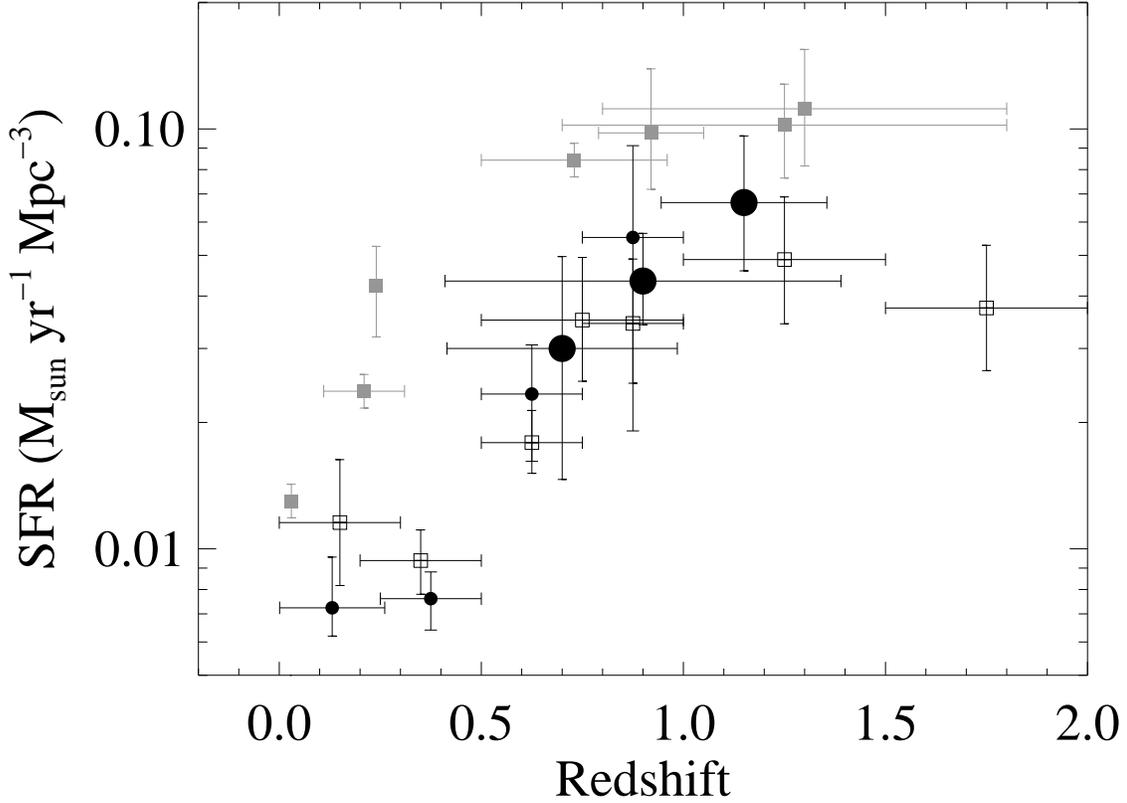}
\caption{Comoving density of star formation as a function of redshift.  Gray filled squares show the values derived
from the \ha\ measurements (in order of increasing redshift) of:  Gallego et al.\ (1995); Tresse \& Maddox (1998); 
Pascual et al.\ (2001); 
Tresse et al.\ (2002); Glazebrook et al.\ (1999); Hopkins et al.\ (2000); Yan et al.\ (1999).  The open
squares are SFR densities inferred from UV continuum measurements (without extinction correction) by:  
Treyer et al.\ (1998); Lilly et al.\ (1996);
Connolly et al.\ (1997).  The small filled circles are the [OII] luminosity density of Hammer et al.\ (1997; converted
to \ha\ by [OII]/\ha =0.45, see Kennicutt 1998) and
Sullivan et al.\ (2000).  The large filled circles are the integrated SPS luminosity function.
The three points are not independent as indicated by the redshift ranges (horizontal bars) -- 
the middle point includes all objects in the survey, while the other points are for the $z<1$\ and $z>1$\ subsamples.
\label{fig: madau} }
\end{figure}

As expected, the SPS data show strong evolution in the density of
star-formation over the range $1.14> z >0.72$.  Although the
uncertainties are involved are large, the evolution observed is
consistent with $(1+z)^4$, the same as that computed by Tresse et al.\ 
(2002) for evolution in \ha. The same exponent is derived when the
other [OII] measurements in Figure \ref{fig: madau}\ are included.
The value agrees with the wavelength-independent exponent of $3.8\pm
0.8$\ for a matter dominated flat Universe (Hogg 2002).  However,
Baldrey et al.\ (2002) set a strong upper limit on the exponent of
$\le 2$\ at $z>1$.

It is apparent from Figure \ref{fig: madau}\ that the [OII]
measurements of the SFR density do not agree with the \ha\ 
measurements any more than with those inferred from the UV continuum.
The clear difference between the \ha\ and UV measurements is usually
attributed to dust extinction (e.g. Yan et al.\ 1999), but at 3727
\AA, the [OII] emission should be subject to less extinction from dust
than the UV continuum.  Furthermore, any difference in
extinction between \ha\ and [OII] should have been included, to first
order, in the calibration of the ratio of the lines by Jansen et al.\ 
(2001), as that ratio was derived empirically from typically metal
rich galaxies.  If more sensitive spectro-photometry were obtained for
the SPS sample, it would be possible to recalculate the inferred 
\ha\ luminosities using reddening information in the Jansen et al.\ fit.
It might be argued that the SPS selection effects
favor more reddened objects, if reddening is greater in more compact
starbursts, perhaps as a result of the greater concentration of gas
and dust.  The CFRS9 measurement, however, shows the same distinction,
and that survey had very different selection criteria.

The disagreement between the [OII] and \ha\ values might be resolved
if both are corrected for extinction.  Gallego et al.\ (2002) find the
extinction correction for [OII] to be a factor of ten on average,
which brings their measurement of the local SFR density into agreement
with the density inferred from \ha\ (Gallego et al.\ 1995).  This
value seems quite large, however, and does not agree with that used in
other studies.  Tresse et al.\ (2002) apply a factor of 2 extinction
correction to \ha\ and only a factor of 3.6 to [OII], similar to
Sullivan et al.\ (2000).

The uncertainties in the measurement of the SFR density are still
large, and the number of objects with both [OII] and \ha\ line fluxes
is still small.  A direct approach to resolving the differences
suggested by Figure 5 is additional multiwavelength study of
consistent samples.  At $z\sim 1$, these measurements require
sensitive near-IR spectroscopy.  Observations of that kind are
possible and becoming more common (e.g. Pettini et al.\ 2001, Teplitz
et al.\ 2000) from the ground, or soon with the refurbished NICMOS.

\acknowledgements

We thank the members of the Space Telescope Imaging Spectrograph
Investigation Definition Team (STIS IDT) for their encouragement of
this project. We acknowledge the contribution of Terry Beck, Ruth Bradley, 
Keith Feggans, Theodore R. Gull, 
Mary E. Kaiser, Philip C. Plait, 
Jennifer L. Sandoval, and Gerard M. Williger.
This research has made use of the NASA/IPAC Extragalactic Database (NED)which is operated
by the Jet Propulsion Laboratory, California Institute of Technology, under contract
with the National Aeronautics and Space Administration.

Funding for this publication was provided by NASA through Proposal number 
HST-AR-08380 submitted to the Space Telescope Science Institute, which is 
operated by the Association of Universities for Research in Astronomy, Inc., under
NASA contract NAS5-26555.  Support was also provided by  
the STIS IDT through the National Optical Astronomical Observatories
and by the Goddard Space Flight Center.  J.R.  was supported by the National Research
Council-GSFC Research Associateship.

\references

\reference{} Calzetti, D. 1997, AJ, 113, 162

\reference{} Cohen, J. G., Hogg, D. W., Blandford, R., Cowie, L. L., Hu, E., Songaila, A., Shopbell, P., \&
Richberg, K.\ 2000, ApJ, 538, 29

\reference{} Connolly, A. J., Szalay, A. S., Dickinson, M., Subbarao, M. U., \& Brunner, R. J.\ 1997, ApJ, 486, L11

\reference{} Cowie, L. L., Songaila, A., Hu, E. M., \& Cohen, J. G.\ 1996, AJ, 112, 839

\reference{} Gallego, J., Zamorano, J., Arag\'{o}n-Salamanca, A., \& Rego, M.\ 1995, ApJ, 455, L1

\reference{} Gallego, J., Garc\'{i}a-Dab\'{o}, C. E., Zamorano, J.,
 Arag\'{o}n-Salamanca, A.,\& Rego, M.\ 2002, ApJ, 570, L1

\reference{} Gardner, J.P., et al.\ 1998, ApJ, 492, L99

\reference{} Glazebrook, K. Blake, C., Economou, F., Lilly, S., \& Colless, M. 1999,\& MNRAS,306, 843

\reference{} Hammer, F., et al.\ 1997, ApJ, 481, 49

\reference{} Hicks, E. K. S., Malkan, M. A., Teplitz, H. I., McCarthy, P. J., \& Yan, L.\ 2002, ApJ in press

\reference{} Hogg, D. W., Cohen, J. G., Blandford, R., Pahre, M. A.\ 1999a, ApJ, 504, 622

\reference{} Hogg, D.W.\ 1999b, astro-ph, 9905116

\reference{} Hogg, D.W.\ 2002, ApJL, submitted; astro-ph, 0105280

\reference{} Hopkins, A. M., Connolly, A. J., \& Szalay, A. S.\ 2000, AJ, 120, 2843

\reference{} Jansen, R. A., Franx, M., Fabricant, D.\ 2001, ApJ, 551, 825

\reference{} Kennicutt, R.C., Jr.\ 1992, ApJ, 338, 310

\reference{} Kennicutt, R. C., Jr.\ 1998, ARA\&A, 36, 189

\reference{} Lilly,S.J., Tresse, L., Le F\`{e}vre, O., Hammer, F., \& Crampton, D.\ 1995, ApJ, 455, 108

\reference{} Lilly, S.J., Le Fevre, O., Hammer, F., \&  Crampton, D.\ 1996, ApJ, 460, L1

\reference{} Madau, P., Ferguson, H.C., Dickinson, M.E., Giavalisco, M., 
        Steidel, C.C., \& Fruchter, A.\ 1996, MNRAS, 283, 1388

\reference{} Madau, P., Pozzetti, L., Dickinson, M.\ 1998, ApJ, 498, 106

\reference{}  McCarthy, P.J., et al.\ 1999, ApJ,520, 548

\reference{} Oke, J. B., et al.\ 1995, PASP, 107, 375

\reference{} Pascual, S., Gallego, J., Arag\'{o}n-Salamanca, A., \&  Zamorano, J.\ 2001, A\&A, 379, 798

\reference{} Pettini, M., Shapley, A.E., Steidel, C.C., Cuby, J.-G., Dickinson, M., Moorwood, A.F.M., 
Adelberger, K.L., \& Giavalisco, M. 2001, ApJ, 554, 981

\reference{} Sullivan, M., Treyer, M. A., Ellis, R. S., Bridges, T. J., Milliard, B., Donas, J.\ 2000, MNRAS, 312, 442

\reference{} Teplitz, H.I, Collins, N.R, Gardner, J.P., Heap, S.R., Hill, R.S., Lindler, D.J., Rhodes, J.,
\& Woodgate, B.E.\ 2002, ApJS submitted 

\reference{} Teplitz, H.I., et al.\ 2000,ApJ, 533, L65 

\reference{} Thompson, R. I., Rieke, M., Schneider, G., Hines, D. C., \& Corbin, M. R.\ 1998, ApJ, 492, L95

\reference{} Tresse, L., \& Maddox, S. J. 1998, ApJ, 495, 691

\reference{} Tresse, L., Maddox, S.J.,  Le Fevre, O., Cuby, J.-G.\ 2002, MNRAS in press; astro-ph, 0111390

\reference{} Treyer, M. A., Ellis, R. S., Milliard, B., Donas, J., \& Bridges, T. J.\ 1998, MNRAS, 300, 303

\reference{} Williams, R.E. et al.\ 1996, AJ, 112,1335

\reference{} Yan, L., McCarthy, P.J., Freudling, W., Teplitz, H.I., Malumuth, E.M., 
\& Weymann, R.J., Malkan, M.A.\ 1999, ApJ, 519, L47 

\clearpage

\begin{deluxetable}{lrr}
\tabletypesize{\scriptsize}
\tablecaption{Incompleteness in the SPS \label{table: incompleteness}}
\tablewidth{0pc}
\tablehead{
\colhead{SNR} &
\colhead{($W_{obs}\sim 100$)} & 
\colhead{($W_{obs}\sim 35$)} \\
\colhead{} &
\colhead{Completeness (\%)} &  
\colhead{Completeness (\%)}
}
\startdata

3 &  39  &  20  \\ 
4 &  56  &  28  \\
5 &  72  &  44  \\
6 &  86  &  58  \\ 
7 &  96  &  69  

 \enddata

\end{deluxetable}

\clearpage

\clearpage

\clearpage

\clearpage

\clearpage

\end{document}